\documentclass{article}
\usepackage{PRIMEarxiv}
\usepackage[utf8]{inputenc}   % allow utf-8 input
\usepackage[T1]{fontenc}      % use 8-bit T1 fonts
\usepackage{url}              % simple URL typesetting
\usepackage{booktabs}         % professional-quality tables
\usepackage{amsfonts}         % blackboard math symbols
\usepackage{nicefrac}         % compact symbols for 1/2, etc.
\usepackage{microtype}        % microtypography
\usepackage{amsmath}
\usepackage{amsthm, amssymb}
\usepackage{fancyhdr}         % header
\usepackage{graphicx}         % graphics
\graphicspath{{media/}}       % images/figures folder
\usepackage{multirow}
\usepackage{listings}
\usepackage{float}
\usepackage{subcaption}
\usepackage{enumitem}
\usepackage{xcolor}
\usepackage[compress]{cite}
\usepackage[colorlinks=true, allcolors=blue]{hyperref}
\usepackage[ruled,vlined,linesnumbered]{algorithm2e}
\IncMargin{1.5em}

\SetKwComment{Comment}{/* }{ */}

\newcommand{\mymatrix}[1]{\ensuremath{\overset{\xrightarrow[\hphantom{#1}]{\text{\scriptsize Subcarrier}}}{#1}\left\downarrow\vphantom{#1}\right.}}

\usepackage{caption}
\usepackage{subcaption}

\title{Through-Wall Detection using Software-Defined Radio based on adaptive Principal Component Analysis}

\author{
Dinuli Naotunna$^{1,2}$, Wenchao Li$^{1}$, Sanka Piyaratna$^{1}$, Phil Wandel$^{1}$\\
$^{1}$Solinnov Pty Ltd\\
$^{2}$Monash University\\
\texttt{\{dinuli.naotunna, wenchao.li\}@solinnov.com.au}\\
\texttt{dnao0001@student.monash.edu}
}

\begin{document}
\maketitle

\begin{abstract}
Through-Wall Detection (TWD) using opportunistic WiFi signals enables non-invasive sensing for security and rescue applications; however many existing approaches rely on controlled access points or specialised hardware. This paper presents a TWD system that extracts Channel State Information (CSI) from ambient WiFi packets using a customised software-defined radio (SDR), Bluebottle, without requiring transmitter control. The key contribution is a spectral-domain scoring mechanism for adaptively selecting motion-relevant principal components from a Principal Component Analysis (PCA) decomposition of the CSI, using Welch power spectral density estimates to quantify each component's signal-to-noise ratio and spectral concentration within the frequency band associated with human motion. The selected components are then analysed using a continuous wavelet transform to robustly identify time-localised motion events. Experimental results demonstrate that the proposed adaptive component-selection method consistently reduces false detections and produces sharper time-frequency energy ridges compared to conventional fixed-component PCA.
\end{abstract}

\section{Introduction}\label{sec:intro}
WiFi devices are ubiquitous and play a central role in modern life by supporting information sharing and communication, with many operating continuously around the clock~\cite{zhang2017wifi,pahlavan2021evolution}. In 2025, approximately 950 million public WiFi hotspots were deployed worldwide, and this number is projected to reach 3.15 billion by 2030~\cite{wifistatistics}. Moreover, to provide reliable coverage, WiFi signals must be capable of penetrating obstacles such as walls, furniture, and glass. Owing to these characteristics, human motion or presence behind walls can be detected by analysing variations in WiFi signals using advanced signal processing techniques, a concept known as Through-Wall Detection (TWD). Recently, TWD has shown potential applications in non-invasive approach for sensing human motion/presence without requiring wearable devices in security and rescue scenarios~\cite{wang2019survey,damodaran2020device}.

Channel State Information (CSI), which captures amplitude and phase variations across subcarriers caused by scattering, fading, and power attenuation\cite{ma2019wifi}, is the most widely used signal for TWD. In practice, CSI is extracted from ambient WiFi packets captured using a Software-Defined Radio (SDR) and then processed to characterise the wireless channel. Early systems often rely on controlled access points, specialised MIMO configurations, or carefully calibrated environments to achieve reliable performance \cite{ali2015keystroke,cao2016wi,wang2017device}. To mitigate noise and multipath effects in CSI, dimensionality-reduction techniques such as Principal Component Analysis (PCA) and its variants have been widely adopted; however, most existing approaches retain a fixed number of leading components or rely on heuristic selection based on explained variance \cite{palipana2016channel, chowdhury2017wihacs,wu2018tw, yang2020pca,showmik2023human}. 
To capture non-stationary motion-induced dynamics in CSI signals, the spectral and time--frequency analysis methods, including short-time Fourier transform (STFT) and wavelet-based transform, have also been employed~\cite{andrews2008enhancing,zhu2025wavelets}. Despite these advances, robust and automatic identification of motion-relevant components in opportunistic sensing scenarios remains an open challenge, particularly in the absence of transmitter control, motivating the adaptive spectral-domain component selection and wavelet-based detection framework proposed in this work.

In this paper, we address a crucial and fundamental problem in TWD, namely, detecting the presence of a human behind wall~\cite{gong2016adaptive,di2018wifi,suraweera2020passive,gu2023wifileaks}. Nonetheless, the algorithm proposed in this paper may be readily generalised to other application scenarios. The main contribution of this paper is an adaptive processing framework for through-wall detection using CSI. Specifically, we introduce an automatic spectral-domain scoring mechanism that enables data-driven selection of motion-relevant principal components in PCA. In the proposed algorithm, the dominant motion-related frequency band is identified from the second principal component of the conventional PCA, then principal components are ranked by a spectral score reflecting how strongly their energy is concentrated within that motion band, and the top-ranked components are selected for detection. Based on this adaptive selection, we further propose a robust detection approach that combines the selected principal components with wavelet-based time-frequency analysis, allowing reliable identification of transient motion events under non-stationary conditions and without requiring control over the WiFi transmitter.

Following this introduction, Section~\ref{sec:formulation} and Section~\ref{sec:conventional} review the background of through-wall detection and provides a brief overview of the PCA-based analysis. Section~\ref{sec:adaptive} presents a signal processing pipeline together with an adaptive principal component selection algorithm. Section~\ref{sec:experiment} reports the experimental results, and Section~\ref{sec:discussion} concludes the paper.

\section{Problem formulation}\label{sec:formulation}

\begin{figure}[htb!]
    \centering
    \includegraphics[width=.7\linewidth]{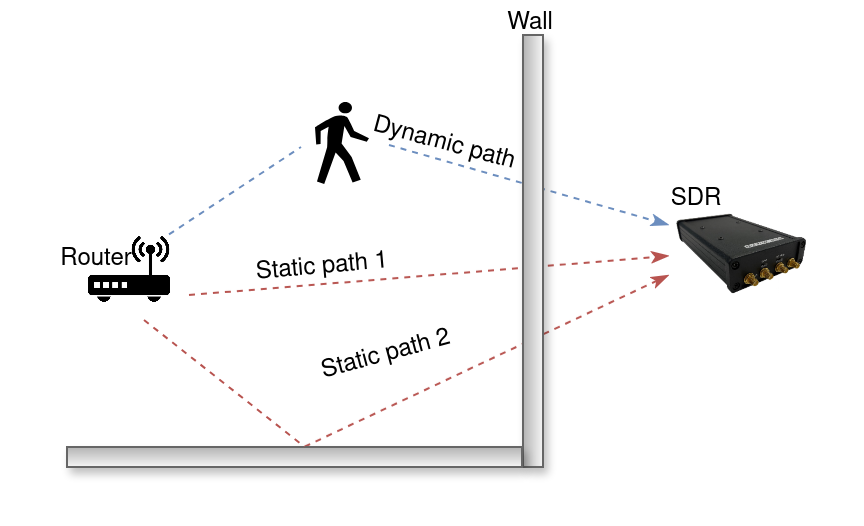}
    \caption{An illustrated example of the TWD.}
    \label{fig:demo}
\end{figure}

CSI characterises how a wireless channel affects transmitted signals. In the frequency domain, let the transmitted signal be $S(f,t)$ and the received signal be $R(f,t)$. Then the received signal can be represented by $R(f,t)=S(f,t)\times H(f,t)$, where $H(f,t)$ is Channel Frequency Response (CFR) for a specific carrier frequency $f$. And the associated Channel Impulse Response (CIR) is $\text{IFFT}(H(f,t))$, where $\text{IFFT}(\cdot)$ is the inverse Fourier transform. In reality, the CFR may be estimated using pilot signal, i.e., $H(f,t)=R(f,t)\times S^*(f,t)$, where $*$ is the conjugation operation. As a result, CSI $H(f_k,t)$ is the sampled version of $H(f,t)$ for $k$-th subcarrier. 
Mathematically, the CSI model can be written as~\cite{ma2019wifi}:
\begin{align}
H(f_k,t)&= e^{-j2\pi\Delta f\, t}
\left(
    \sum_{m\in P_s} a_m(f_k,t)
    e^{-j2\pi \frac{d_m f_k}{c}}
+ \sum_{n\in P_d} a_n(f_k,t)
    e^{-j2\pi \frac{d_n f_k}{c}}
\right)\\
&\triangleq e^{-j2\pi\Delta f\, t}\left(H_s(f_k,t)+H_d(f_k,t)\right)
\end{align}
where $H_s(f_k,t)\triangleq\sum_{m\in P_s} a_m(f_k,t)
    e^{-j2\pi \frac{d_m f_k}{c}}$ and $H_d(f_k,t)\triangleq\sum_{n\in P_d} a_n(f_k)
    e^{-j2\pi \frac{d_n f_k}{c}}$ are the CSI of static and dynamic paths, $P_s$ and $P_d$ are the collections of these two types of paths, $a_m$ and $a_n$ are the amplitudes, $d_m$ and $d_n$ are the path lengths, $\Delta f$ is the carrier frequency difference between the sender and the receiver,  $c$ is the speed of light. An example of static and dynamic paths is shown in Fig.~\ref{fig:demo}.

Suppose that the human's velocity is $v$~m/s, the associated Doppler frequency can be written as~\cite{qian2017widar}
\begin{align}
    f_{D,n,k}(t) = \frac{1}{\lambda_k}\frac{\text{d}}{\text{d}t}d_n(t)
\end{align}
where $f_{D,n,k}$ is the Doppler frequency of $n\in P_d$ path for $k$ sub-carrier,

Consider a simplified scenario, suppose that there exists one dynamic path induced by human movement, then the amplitude of CSI can be written as follows:
\begin{align}
|H(f_k,t)|^2=&|H_s(f_k,t)+H_d(f_k,t)|\notag\\
=&A_s(t)^2+A(t)_d^2+A(t)_dA_h(t)(\exp^{j(\phi_d-\phi_s)}+\exp^{-j(\phi_d-\phi_s)})\notag\\
=&A_s(t)^2+A_d(t)^2+2A_s(t)A_d(t)\cos(2\pi f_{D,n,k}(t)-\phi_s)\label{simplified_model}
\end{align}

From \eqref{simplified_model}, it can be observed that the Doppler frequency explicitly appears in the amplitude of the CSI. In an ideal scenario, the velocity of a human target could be estimated by extracting this Doppler component. However, in practice, this is non-trivial. When a human approaches the direct path (DP) between the router and receiver, the received power $|H(f_k,t)|^2$ is dominated by the strong DP component $A_s$. As a result, the Doppler-induced oscillatory term is masked by the large DC component and further overwhelmed by noise, making reliable Doppler detection difficult. On the other hand, \cite{niu2018fresnel,zhang2019towards,liu2023towards} have shown that the CSI undergoes deep fading when an obstacle enters the first Fresnel zone. This deep fading causes rapid variations in the local frequency of the CSI amplitude, making direct Doppler frequency estimation more difficult; nevertheless, it motivates indirect detection strategies that exploit motion-induced spectral structure and temporal variability instead of explicit Doppler estimation. In this paper, this deep fading effect is used as feature for the detection of human presence. 

\section{Conventional detection algorithm}\label{sec:conventional}

\subsection{Data collection and preparation}
Suppose that the SDR collects $T$~second of CSI data with sampling rate $f_s$ for $k$-th sub-carrier, $k=-K,\cdots,K$. Therefore, the total number of samples/packets per frame is $N=f_sT$. It should be noted that the sampling rate $f_s$ is the number of packets captured by SDR per second and should not be confused with the sampling rate of ADC. As a result, for $i$-th frame, $i=1,\cdots$, CSI data can be arranged into matrix $\mathbf{H}_i\in\mathbb{C}^{N\times (2K+1)}$ as follows:
\begin{align}
\mathbf{H}_i=\mymatrix{\begin{bmatrix}
CSI_{1,-K,i} & CSI_{1,-K+1,i} & \cdots & CSI_{1,K-1,i} & CSI_{1,K,i} \\
CSI_{2,-K,i} & CSI_{2,-K+1,i} & \cdots & CSI_{2,K-1,i} & CSI_{2,K,i} \\
\vdots & \vdots &  \vdots & \vdots & \vdots \\
CSI_{N-1,-K,i} & CSI_{N-1,-K+1,i} & \cdots & CSI_{N-1,K-1,i} & CSI_{N-1,K,i} \\
CSI_{N,-K,i}  & CSI_{N,-K+1,i}  & \cdots & CSI_{N,K-1,i}  & CSI_{N,K,i} 
\end{bmatrix} }\text{\scriptsize Time}\label{H}
\end{align}
Additionally, we denote the entries of matrix $|\mathbf{H}_i|$ by $\left\{|CSI_{n,k,i}|^2\right\}$, for $n=1,\cdots,N$ and $k=-K,\cdots,K$, i.e. the point-wise amplitude of $\mathbf{H}_i$.

\subsection{Digital filter}

To suppress out-of-band noise while preserving motion-related CSI variations, each CSI time series was filtered using a digital Butterworth filter. In the implementation, the filter coefficients were designed using SciPy's \texttt{signal.butter}, and zero-phase filtering was applied using \texttt{signal.filtfilt} to avoid phase distortion. Based on prior observations that motion-induced CSI fluctuations are concentrated mainly between 10~Hz and 35~Hz, the passband was chosen as $[10,35]$~Hz. The filter order was set to 4, and filtering was applied independently to each subcarrier time series.

\subsection{Principal Component Analysis}\label{sec: conventionalPCA}

Principal Component Analysis (PCA) is an orthogonal linear transformation that de-correlates multivariate data and projects it onto a new coordinate system such that the greatest variance, i.e. the most important component, in the data lies on the first coordinate while the least variance, i.e. the least important component, in the data tends to lie on the last coordinate~\cite{hargrove2008principal}. This allows dimensionality reduction by retaining only a subset of components. 

Specifically, for the $i$-th frame, PCA is performed on matrix $|\mathbf{H}|_i \in \mathbb{R}^{N \times (2K+1)}$ defined in~\eqref{H}. Since the following algorithm can be applied on each frame $i=1,2,\cdots$ directly, therefore the subscript $i$ in $\mathbf{H}_i$ and $|\mathbf{H}_i|$ will be dropped for the sake of simplification. 

To remove the static component of each subcarrier, the mean of each column of $\mathbf{X} \triangleq |\mathbf{H}|$ is computed as
\begin{equation}
\bar{\mathbf{x}} = \frac{1}{N} \sum_{n=1}^{N} \mathbf{x}_{n}
\end{equation}
where $\mathbf{x}_{n}\in \mathbb{R}^{1 \times M}$ is the $n$-th column of $\mathbf{X}$. Then, the mean-centered data matrix is then given by
\begin{equation}
\mathbf{X}_c = \mathbf{X} - \mathbf{1}_N\bar{\mathbf{x}}
\end{equation}
where $\mathbf{1}_N$ is a $N$-dimensional column vector of ones.
The sample covariance matrix of the centered CSI amplitudes is computed as
\begin{equation}
\mathbf{\Sigma} = \frac{1}{N-1} \mathbf{X}_c^T\mathbf{X}_c \in \mathbb{R}^{M \times M}
\end{equation}
PCA is then performed via eigen-decomposition of the covariance matrix:
\begin{equation}
\mathbf{\Sigma} = \mathbf{V}\boldsymbol{\Lambda}\mathbf{V}^T
\end{equation}
where $\mathbf{V} = [\mathbf{v}_1, \mathbf{v}_2, \ldots, \mathbf{v}_M]\in \mathbb{R}^{N \times M}$ contains the orthonormal eigenvectors and $\boldsymbol{\Lambda} = \mathrm{diag}(\lambda_1, \lambda_2, \ldots, \lambda_M)$ is a diagonal matrix of eigenvalues ordered as $\lambda_1 \ge \lambda_2 \ge \cdots \ge \lambda_M \ge 0$. Each eigenvalue $\lambda_i$ represents the variance explained by the corresponding principal component.

The principal component time series are obtained by projecting the centered data onto the principal directions:
\begin{equation}
\mathbf{Z} = \mathbf{X}_c \mathbf{V}
\end{equation}
and $i$-th principal component is given by
\begin{equation}
\mathbf{z}_i = \mathbf{X}_c \mathbf{v}_i\text{ for }i=1,\cdots,M
\end{equation}

The leading principal components capture the dominant temporal variations shared across subcarriers, while higher-order components primarily correspond to noise and subcarrier-specific fluctuations. Therefore, in conventional PCA, a fixed number of consecutive principal components is typically retained starting from the second component, under the assumption that motion-related channel variations are captured by components whose power decreases progressively. In practice, however, this assumption may not hold in the presence of broadband noise and interference. This limitation motivates the adaptive approach presented in Section~\ref{sec:adaptive}.

\subsection{Wavelet transform}

To further analyse the non-stationary structure of the extracted motion trace (or selected PCA-derived signal), the wavelet-based time--frequency analysis is usually applied. Unlike Fourier-based methods that assume stationarity over the analysis window, wavelet transform provides joint time-scale localisation, making them well suited for signals whose spectral content evolves over time, as is typical in motion-driven CSI measurements. In particular, we use the continuous wavelet transform (CWT) to obtain highly interpretable time--frequency representation (scalogram), in which transient or localized events appear as concentrated regions of high energy. The CWT of a signal $x(t)$ can be formally described by:
\begin{equation}
W_x(a,b)=\frac{1}{\sqrt{|a|}}\int_{-\infty}^{\infty} x(t)\,\psi^{*}\!\left(\frac{t-b}{a}\right)\,dt,
\end{equation}
where $x(t)$ is signal, $a>0$ is the scale parameter, $b$ is the time shift, $\psi(\cdot)$ is the chosen mother wavelet, and $(\cdot)^{*}$ denotes complex conjugation. The corresponding scalogram $|W_x(a,b)|$ (often mapped from scale to pseudo-frequency) highlights time-localized energy bursts, which in our application manifest as visually salient and bright vertical ridges indicating candidate detection events, i.e. the presence of human. This is actually due to the fact that CWT-based scalograms are preferred for identifying transient and multi-scale structures in non-stationary signals~\cite{torrence1998practical,mallat2009wavelet}. On the other hand, for a given signal sampled at rate $f_s$, the dyadic scale levels are constructed from the frequency range of interest by computing
\[
\ell = \log_2\!\left(\frac{f_s/2}{f}\right),
\]
where $f$ denotes the specified frequency bounds, and then generating $L$ logarithmically spaced scales over this range, resulting to log-uniform spaced pseudo-frequencies. Eventually, this procedure yields a time--frequency representation with localisation of transient motion-induced events across the frequency band of interest.

\subsection{Detection}

To reduce computational complexity, detection of human presence can be implemented using the Constant False Alarm Rate (CFAR) method~\cite{farina1986review}. Although the wavelet scalogram is a two-dimensional matrix, only the time axis is of interest in this study. Therefore, the scalogram can be transformed into a one-dimensional vector representing the total energy within each time interval, i.e.,
\begin{align}
    v_j = \sum_{i=1}^LW_x(f_i, t_j)\label{detection}
\end{align}
where $L$ is the number of scales used in the wavelet. Therefore, we have $\mathbf{v}=[v_1,\cdots,v_M]$ and $M$ is the number of time grid. As a result, the CFAR can be applied.

\section{Adaptive Principal Component Analysis}\label{sec:adaptive}

\subsection{Introduction}
In Section~\ref{sec:conventional}, it is shown that the selection of principal components significantly affects the signal-to-noise ratio (SNR) of the resulting spectrogram used for detection. Conventionally, the first principal component is always discarded, as it is often assumed to capture dominant static or noise-related variations, and a fixed number of subsequent consecutive components are selected~\cite{wang2017device,ali2015keystroke,cao2016wi, wu2018tw, showmik2023human}. For example, the first $5\sim 20$ principal components are used for detection in~\cite{wu2018tw} while only the 2nd and 3rd components are used in~\cite{showmik2023human}. Theoretically, motion-related information is distributed across all principal components, with the explained variance decreasing according to component order. However, in practice, the energy distribution among components may be uneven~\cite{palipana2016channel} and depends on the environment because of noise and environmental complexity. As a result, the selected components may not contain sufficient motion-related energy to be effectively revealed by the wavelet transform, leading to unreliable detection due to low SNR. Moreover, using a fixed selection of principal components does not generalise well across different environments, as channel conditions and multipath characteristics vary, altering the energy allocation among components.

\begin{figure}[htb!]
    \centering
    \includegraphics[width=0.8\linewidth]{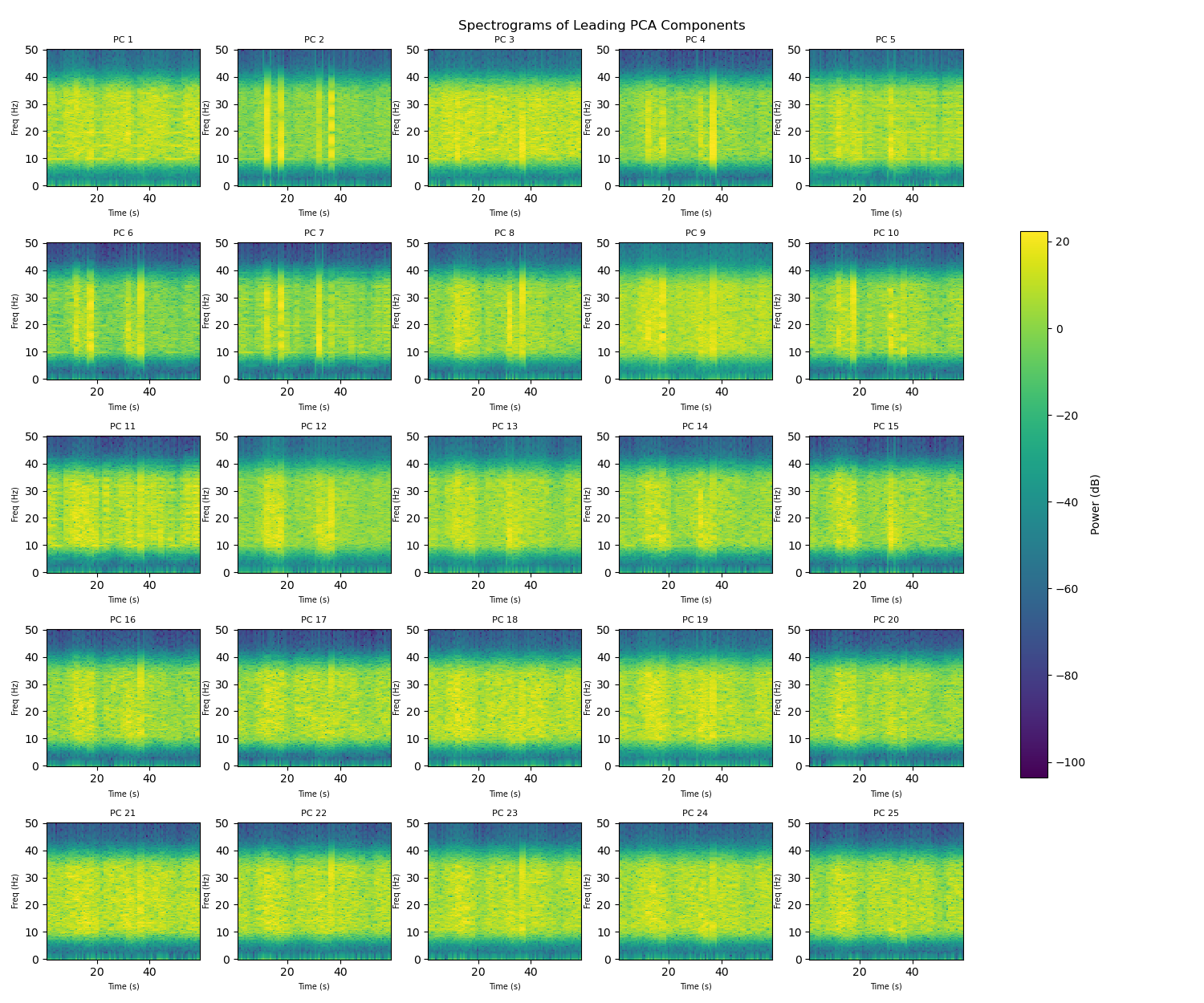}
    \caption{The first $25$ PCA components showing the energies of movements.}
    \label{fig:example}
\end{figure}

An example is used to illustrate the scenario. In this example, the router and SDR are placed in two distinct office rooms dividing by a wall, where the distance is approximately $8$~m, see Office 1 in Fig.~\ref{fig:demo6}. A human crosses the line of router and SDR at $12$s, $18$s, $32$s and $38$s separately during the total data collection duration, $60$s. The spectrogram of first $25$ PCA components are plotted in Fig.~\ref{fig:example}. It can be seen that the movements are clear in components $\{2, 4, 6, 7, 8\}$, and obscure in components $\{3, 5, 9, 11\}$.

Therefore, it is desirable to develop an algorithm that can adaptively select the appropriate components from the PCA output, thereby improving the overall SNR while making the system less sensitive to environmental variations and noise. Inspired by~\cite{andrews2008enhancing}, a Power Spectral Density (PSD)  based method is proposed to mitigate this problem. 

\subsection{Power Spectral Density analysis for PCA components selection}

PSD measures how the power (or variance) of a signal is distributed across frequencies. There are a series of methods to estimate the PSD~\cite{kay2005spectrum,welch2003use,han2023welch}, such as FFT based method, short-time FFT based method and Welch method.  Compared to a classical periodogram based on FFT, which yields high-variance spectral estimates, Welch’s approach improves robustness by partitioning the signal into overlapping segments, applying windowing to reduce spectral leakage, and averaging the resulting periodograms~\cite{zhu2025wavelets}. This variance reduction is particularly important in the present setting, as the principal component signals derived from CSI amplitudes are inherently non-stationary due to time-varying propagation conditions, human motion, and environmental dynamics. By assuming approximate stationary over short segments rather than over the entire observation window, Welch’s method provides a stable and reliable estimate of the spectral energy distribution without imposing restrictive parametric assumptions \cite{chowdhury2017wihacs,andrews2008enhancing}. Therefore, in this paper, the Welch's PSD is used to quantify the motion relevance of each principal component. 

In practice, the first principal component typically captures dominant static background structure rather than motion-induced dynamics, and is therefore discarded. Among the remaining components, the second principal component is the most reliable reference for characterising motion-related spectral content. Because it captures the largest share of variance after the static background has been removed, it has the highest SNR of the motion-relevant components and is therefore least affected by noise contamination. This makes its dominant spectral peak a stable and consistent indicator of the frequency band in which motion energy is concentrated.
Accordingly, the second principal component is used solely to define this reference frequency band. Specifically, a fixed window $\mathcal{B}=[f_c-\Delta f/2,\;f_c+\Delta f/2]$ centred on this peak is applied uniformly across all principal components, where $f_c$ denotes the reference frequency and $\Delta f$ the bandwidth. In our experiments, the bandwidth was set to $\Delta f = 7$~Hz. This value was chosen empirically to provide adequate coverage of the dominant motion-related spectral peaks observed in the Welch's PSD of the principal components, while avoiding an excessively broad band that would include unrelated spectral energy.

For the $k$-th principal component, $k=3,4,\cdots$, the in-band spectral power is computed as
\begin{equation}
P_k^{\mathrm{in}} = \int_{\mathcal{B}} \hat{P}_k(f)\,df,
\end{equation}
and the total spectral power as
\begin{equation}
P_k^{\mathrm{tot}} = \int_{\mathcal{F}} \hat{P}_k(f)\,df,
\end{equation}
where $\mathcal{F}$ denotes the effective frequency span of the PSD. The band ratio is then defined as
\begin{equation}
\text{BR}_k = \frac{P_k^{\mathrm{in}}}{P_k^{\mathrm{tot}}},
\end{equation}
which measures the degree of spectral energy concentration within the motion-related frequency band.

To further assess robustness against broadband noise, the mean in-band and out-of-band spectral levels are computed as
\begin{equation}
\mu_k^{\mathrm{sig}} = \frac{P_k^{\mathrm{in}}}{|\mathcal{B}|},
\qquad
\mu_k^{\mathrm{noise}} =
\frac{P_k^{\mathrm{tot}} - P_k^{\mathrm{in}}}{|\mathcal{F}| - |\mathcal{B}|},
\end{equation}
where $|\mathcal{B}|$ and $|\mathcal{F}|$ denote the bandwidths of the in-band region and the full PSD support, respectively. A SNR–like ratio is then defined as
\begin{equation}
\text{SNR}_k =
\frac{\mu_k^{\mathrm{sig}}}{\mu_k^{\mathrm{noise}} + \epsilon},
\end{equation}
with a small constant $\epsilon>0$ introduced for numerical stability \cite{yang2020pca,coluccia2020cfar}. $\mathrm{SNR}_k$ is termed as equivalent SNR in this paper.

Then the motion relevance of the $k$-th principal component is quantified by a scalar score
\begin{equation}
s_k = \text{BR}_k \cdot \text{SNR}_k ,
\label{eq:score}
\end{equation}
where higher scores indicate components that exhibit both strong spectral localisation and robustness against broadband noise. Principal components with the highest scores are therefore selected as those most strongly associated with motion-induced channel variations.

As a result, the set of selected components, denoted by $\mathcal{I}$, is defined as the set of indices corresponding to the top-$m$ highest-scoring principal components, i.e.
\begin{equation}
\mathcal{I} = \arg \mathrm{top}_m\left\{s_k,\; k=2,\cdots,2K+1\right\}
\label{eq:topm}
\end{equation}
where operator $\arg \mathrm{top}_m$ returns the indices of the $m$ largest elements in the set. In this work, $m=5$.

The motion detection algorithm using adaptive PCA is listed in Algorithm~\ref{alg:one} while the adaptive PCA is listed in Algorithm~\ref{alg:two}, where the \textbf{ConventionalPCA}($\cdot$) is the  conventional PCA as shown in Section~\ref{sec: conventionalPCA}.

\begin{algorithm}
\caption{Presence detection algorithm}\label{alg:one}
\KwData{$\mathbf{H}_i$, $m$} 
$\mathbf{B}\gets\mathbf{0}_{N\times(2K+1)}$\;
\KwResult{Time}
\For{$n=1:N$}{
$\mathbf{B}[n,:]=\textbf{LowPassFilter}(\mathbf{H}_i[n,:])$\Comment*[r]{Low pass filter}
}
$\mathcal{I}, \mathbf{C}\gets\textbf{AdaptivePCA}(\mathbf{B},m)$\Comment*[r]{Adaptive PCA}
$\mathbf{v}\gets\frac{1}{\#\mathcal{I}}\sum_{n\in\mathcal{I}}\mathbf{C}[n,:]$\Comment*[r]{Averaging the components}
$\mathbf{W}\gets\textbf{CWT}(\mathbf{v})$\Comment*[r]{Continuous wavelet transform}
$\mathbf{w}\gets\frac{1}{L}\sum_{l=1}^L\mathbf{W}[l,:]$\Comment*[r]{Convert the scalogram to 1D}
$\text{Time}\gets\textbf{CFAR}(\mathbf{w})$\Comment*[r]{1D CFAR}
\end{algorithm}

\begin{algorithm}
\caption{\textbf{AdaptivePCA} - Adaptive PCA algorithm}\label{alg:two}
\KwData{$\mathbf{B}, m$} 
\KwResult{$\mathbf{C}, \mathcal{I}$}
$\mathbf{C}\gets\textbf{ConventionalPCA}(\mathbf{B})$\;
\For{$k=1:K$}{
$\mathbf{A}[k,:]\gets\textbf{WelchAlgorithm}(\mathbf{C}[k,:])$\Comment*[r]{Spectral calculation}
}
$\mathcal{B}\gets[f_c-\Delta f/2,\;f_c+\Delta f/2]$\;
$\mathcal{F}\gets[f_c-f_s/2,\;f_c+ f_s/2]$\;
\For{$k=2:2K+1$}{
$\text{BR}_k\gets\frac{P_k^{\mathrm{in}}}{P_k^{\mathrm{tot}}}$\Comment*[r]{Calculate band ratio}
$\text{SNR}_k\gets
\frac{\mu_k^{\mathrm{sig}}}{\mu_k^{\mathrm{noise}} + \epsilon}$\Comment*[r]{Calculate equivalent SNR}
$s[k]\gets\text{BR}_k \cdot \text{SNR}_k$\;
}
$\mathcal{I}\gets\arg \mathrm{top}_m\left\{s_k,\; k=2,\cdots,2K+1\right\}$\Comment*[r]{top-$m$ highest-scoring components}
\end{algorithm}

\section{Experimental validation}\label{sec:experiment}
\subsection{Experimental setup and data collection}
\begin{figure}[htb!]
    \centering
    \includegraphics[width=0.7\textwidth]{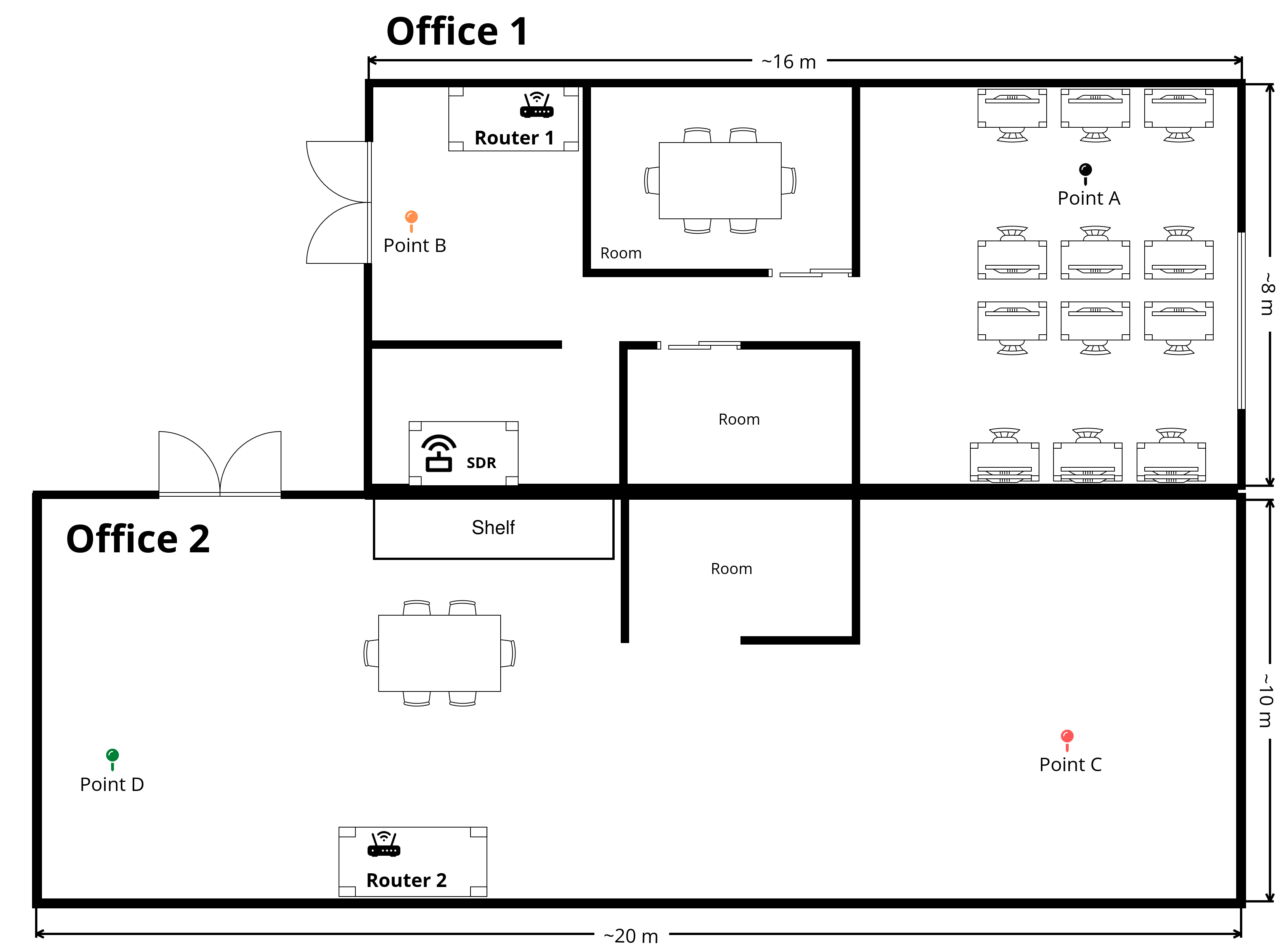}
    \caption{Experimental environment (not to scale). }
    \label{fig:demo6}
\end{figure}

In this section, experiments are conducted to evaluate the performance of the proposed opportunistic WiFi-based TWD system using SDR Bluebottle~\cite{solinnovbb}, a customised SDR platform designed and manufactured by Solinnov for a wide range of wireless sensing and measurement applications. Measurements were collected inside Solinnov’s Melbourne office using two commercial WiFi routers as transmitters and a Bluebottle as the receiver. In the experiment, it was configured to operate as a passive WiFi receiver, capturing ambient packets and extracting CSI without requiring coordination with the transmitter. This allows opportunistic CSI collection from existing WiFi traffic in the experimental environment. Fig.~\ref{fig:demo6} illustrates the office layout, the placement of the routers and Bluebottle, as well as the start/end points of the walking trajectories. Data were extracted from all received packets containing usable CSI estimates, including beacon, probe, and other management or data frames. Because native packets were irregular and raw traffic did not provide a uniformly sampled time series suitable for spectral analysis, the CSI sequence was interpolated and resampled to an effective sampling rate of $f_s=100$~Hz. The processed data were then analysed in frame-wise observation windows, where the duration of the frame is denoted by $T$ and may vary between scenarios. Accordingly, the number of samples in the $i$-th frame is
$N = f_s T$. 
The corresponding CSI measurements for frame $i$ are arranged into the matrix $\mathbf{H}_i \in \mathbb{C}^{N \times (2K+1)}$. Additionally, CWT was implemented using the Morlet wavelet as the mother wavelet~\cite{cohen2019better}.

Multiple experiments were performed in which a volunteer walked repeatedly between two points resulting in multiple crossings of the DP between the router and Bluebottle. To capture realistic variability, walking speed and dwell time were varied across trials. The following section presents representative results from three trials and compares conventional PCA (cPCA) components selection with the proposed adaptive PCA (aPCA) components selection.

In particular, multiple experiments were conducted in two offices, whose layouts and the positions of the SDR and routers are shown in Fig.~\ref{fig:demo6}. The experiments were divided into two sets:
\begin{enumerate}
    \item[S1:] In the first set of experiments, the volunteer walked between Point A and B in Office 1, crossing the DP between Router 1 and the SDR. During this process, the SDR collected the signals transmitted by Router 1. Both SDR and Router 1 were located in Office 1.
    \item[S2:] In the second set of experiments, the volunteer walked between Point C and D in Office 2, crossing the DP between Router 2 and the SDR. During this process, the SDR collected the signals transmitted by Router 2. The SDR was located in Office 1, while Router 2 was located in Office 2. S2 represents a more challenging environment due to the greater router-to-SDR distance and the presence of more complex obstacles along the propagation path.
\end{enumerate}

\subsection{Experimental results}

In this section, the performance of the TWD using cPCA is compared with the proposed aPCA under four representative scenarios, i.e. 1. no direct path (DP) crossing; 2. two DP crossings; 3. four DP crossings and; 4. over separate offices. In the cPCA, the fixed components $2$ to $6$ are selected by default, whereas the proposed method adaptively selects the five highest-scoring components based on the spectral-domain scoring framework described in Section~\ref{sec:adaptive}.

\subsubsection{No DP crossing (S1)}\label{s1_1}
\begin{figure}[htb!]
    \centering
    \begin{subfigure}{0.48\linewidth}
        \centering
        \includegraphics[width=\linewidth]{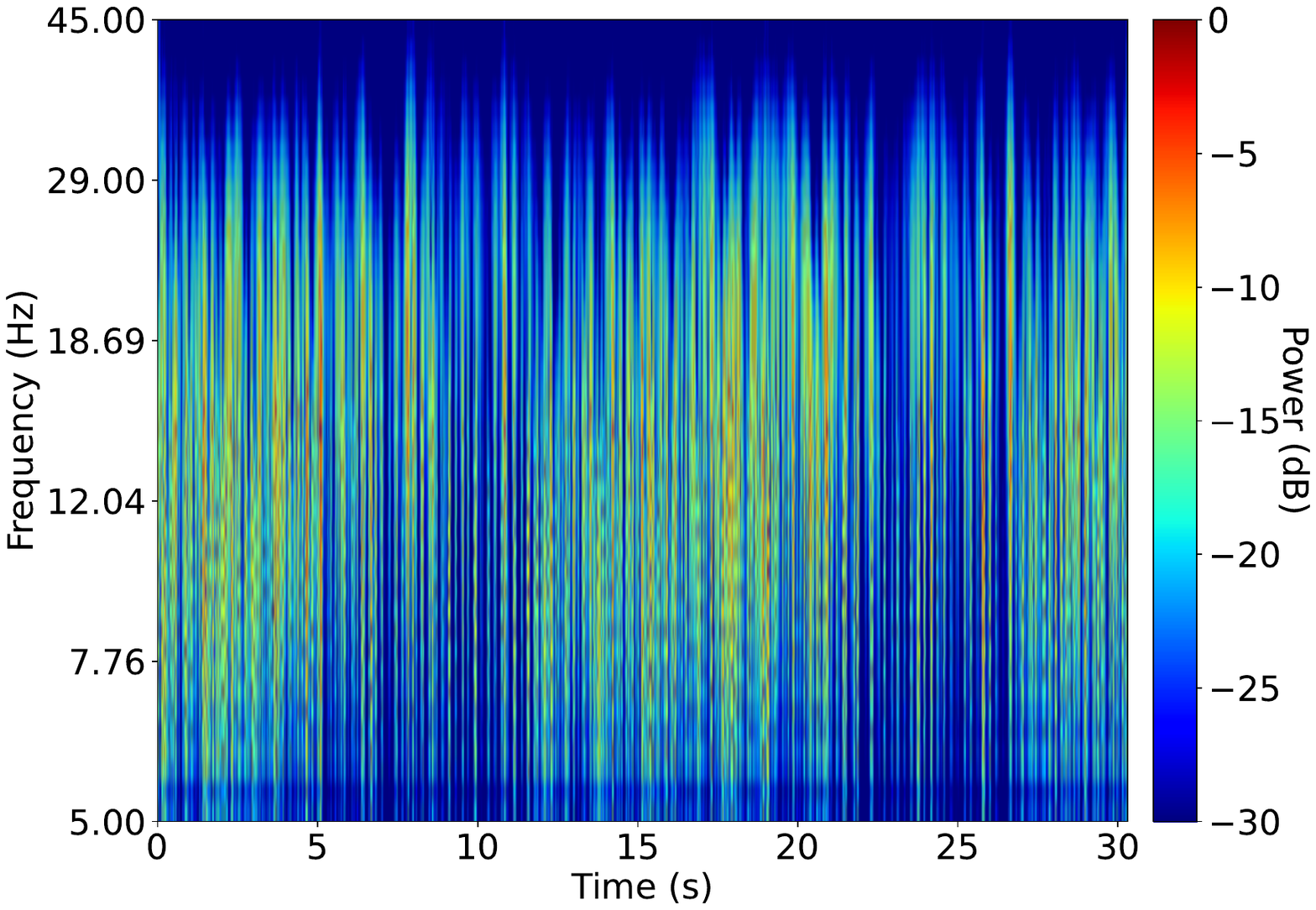}
        \caption{cPCA.}
        \label{fig:cwt_baseline3}
    \end{subfigure}
    \hfill
    \begin{subfigure}{0.48\linewidth}
        \centering
        \includegraphics[width=\linewidth]{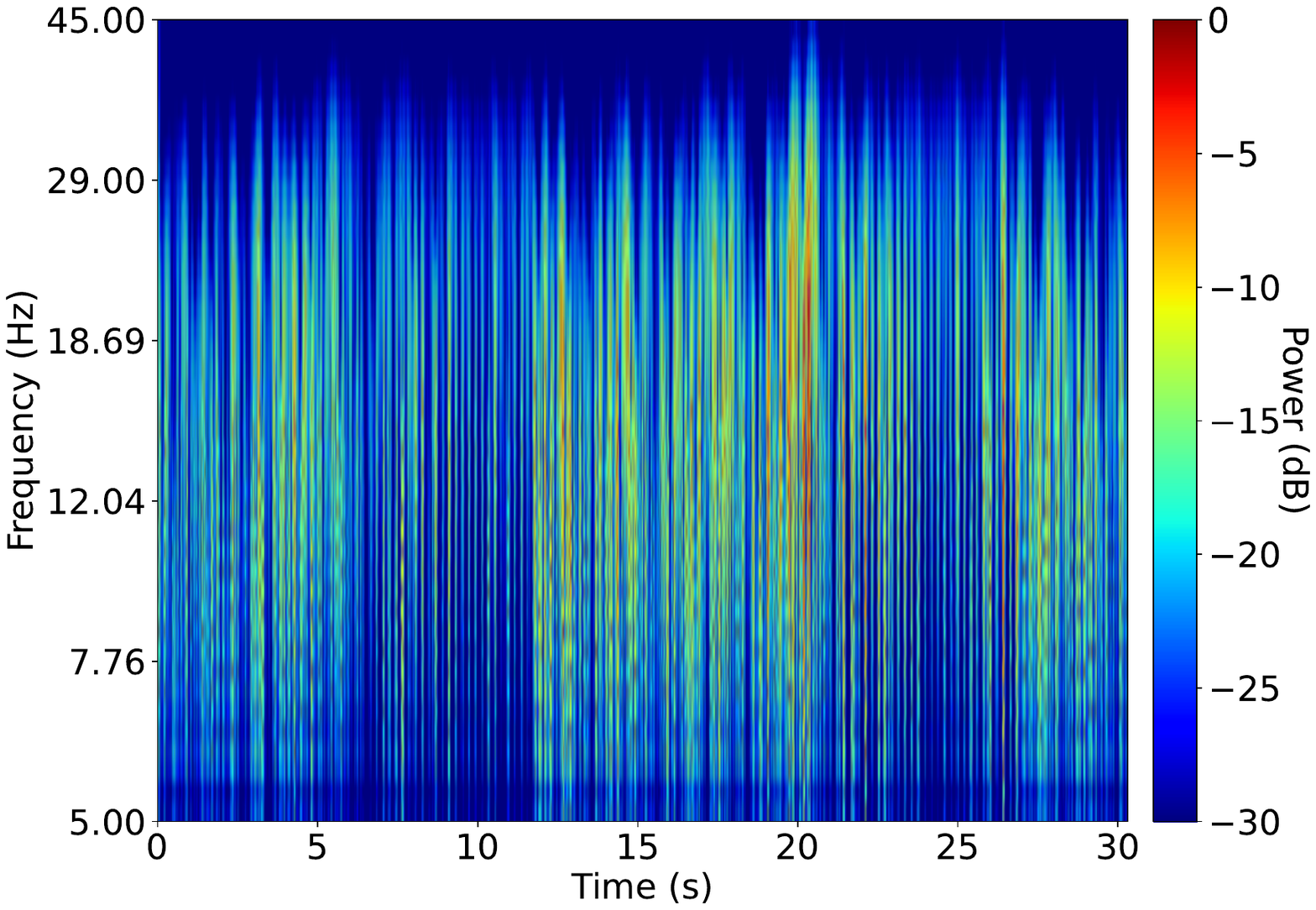}
        \caption{aPCA.}
        \label{fig:cwt_adaptive3}
    \end{subfigure}
    \caption{CWT scalograms for the no DP crossing scenario.
    }
    \label{fig:no_crossings}
\end{figure}

This scenario represents an experiment in which no DP crossing occur and the environment remains largely static. As shown in Fig.~\ref{fig:no_crossings}, the approaches using cPCA and aPCA produce qualitatively similar results, as expected in the absence of motion-induced perturbations. The selections of components and the associated scores $s_k$ for both methods are listed in Table~\ref{tab:pca_scores_no_crossing}.
This scenario confirms that both of conventional and adaptive selection strategy can maintain robustness and do not introduce artificial detections when no motion is present.
\begin{table}[htb!]
\centering
\renewcommand{\arraystretch}{1.2}
\setlength{\tabcolsep}{8pt}
\footnotesize
\begin{tabular}{cccccc}  
\toprule
 \multicolumn{2}{c}{cPCA} & &
\multicolumn{2}{c}{aPCA} \\ 
\cmidrule{1-2} \cmidrule{4-5}
{\centering Components ($k$)} & {Score $s_k$} & & {Components ($k$)} & {Score $s_k$}  \\
\midrule
2 & 0.625 && 1 & 0.861 \\
3 & 0.787 && 5 & 0.787 \\
4 & 0.861 && 4 & 0.750 \\
5 & 0.721 && 15 & 0.721 \\
6 & 0.556 && 22 & 0.709 \\
\bottomrule
\end{tabular}

\caption{Comparison between the cPCA and aPCA components selection for the no DP crossing scenario.}\label{tab:pca_scores_no_crossing}
\end{table}

    \subsubsection{Two DP crossings (S1)}\label{s1_2}
\begin{figure}[htb!]
    \centering
    \begin{subfigure}{0.48\linewidth}
        \centering
        \includegraphics[width=\linewidth]{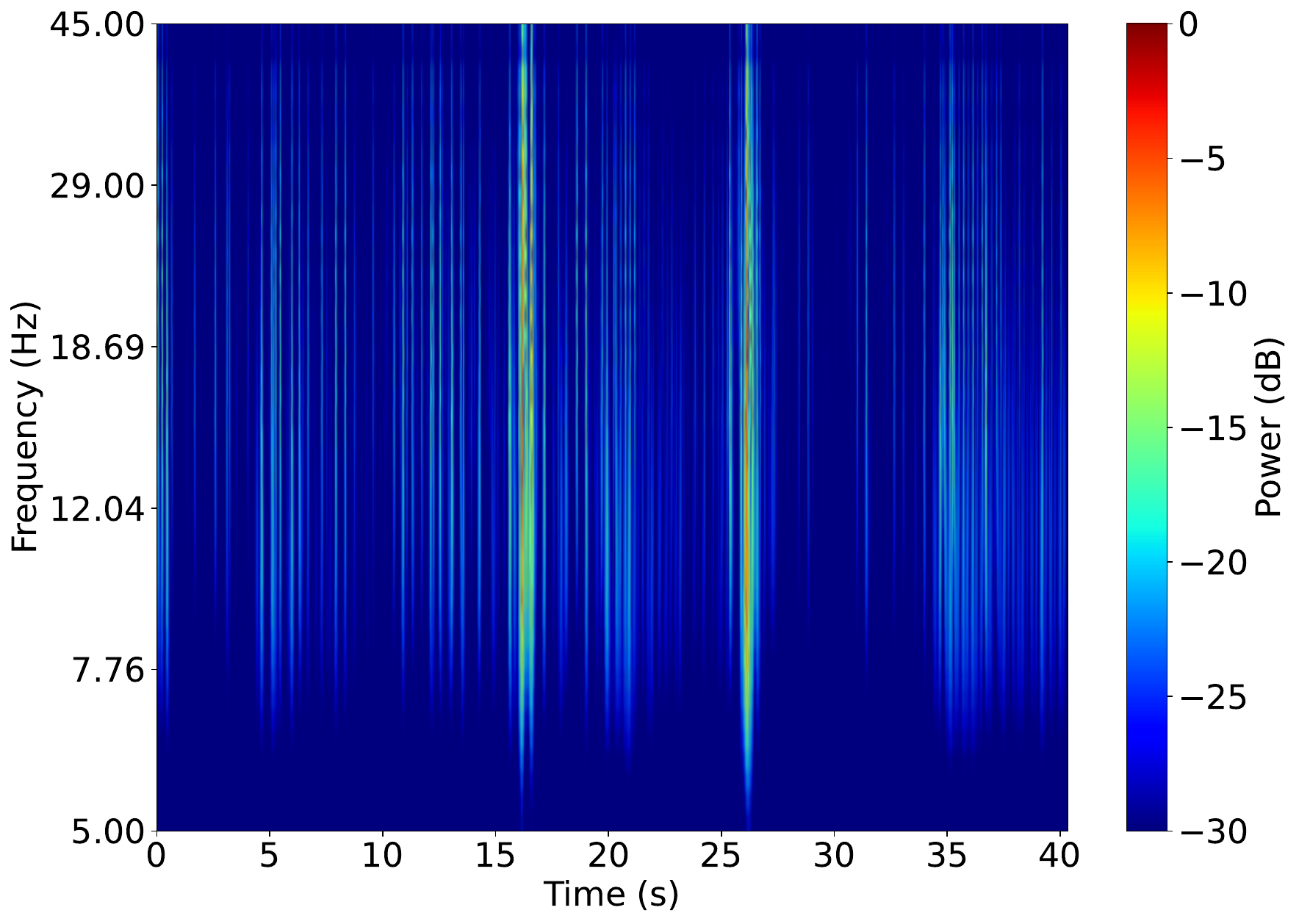}
        \caption{cPCA.}
        \label{fig:cwt_baseline1}
    \end{subfigure}
    \hfill
    \begin{subfigure}{0.48\linewidth}
        \centering
        \includegraphics[width=\linewidth]{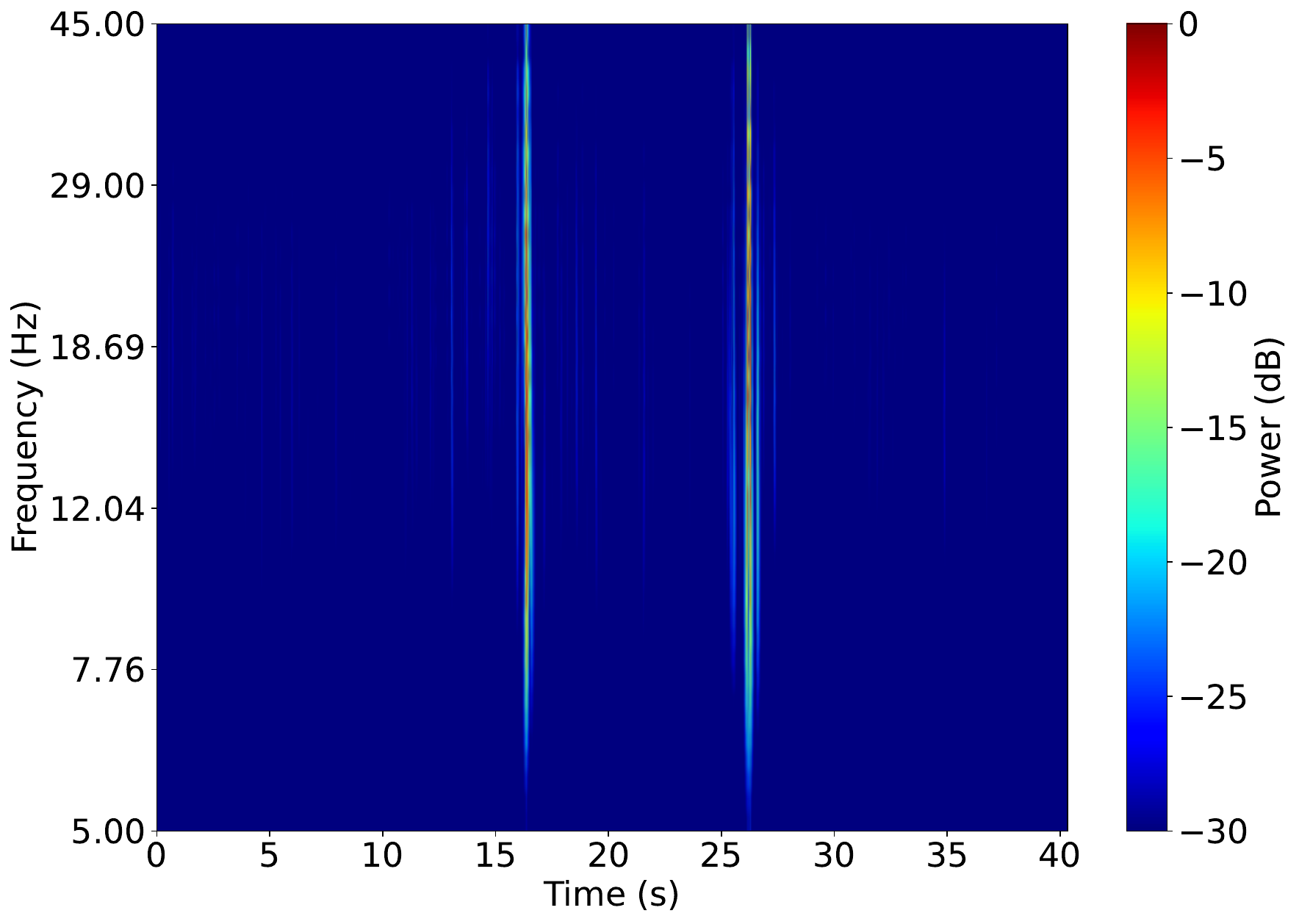}
        \caption{aPCA.}
        \label{fig:cwt_adaptive1_1}
    \end{subfigure}\\[0.2in]
    \begin{subfigure}{0.48\linewidth}
        \centering
        \includegraphics[width=\linewidth]{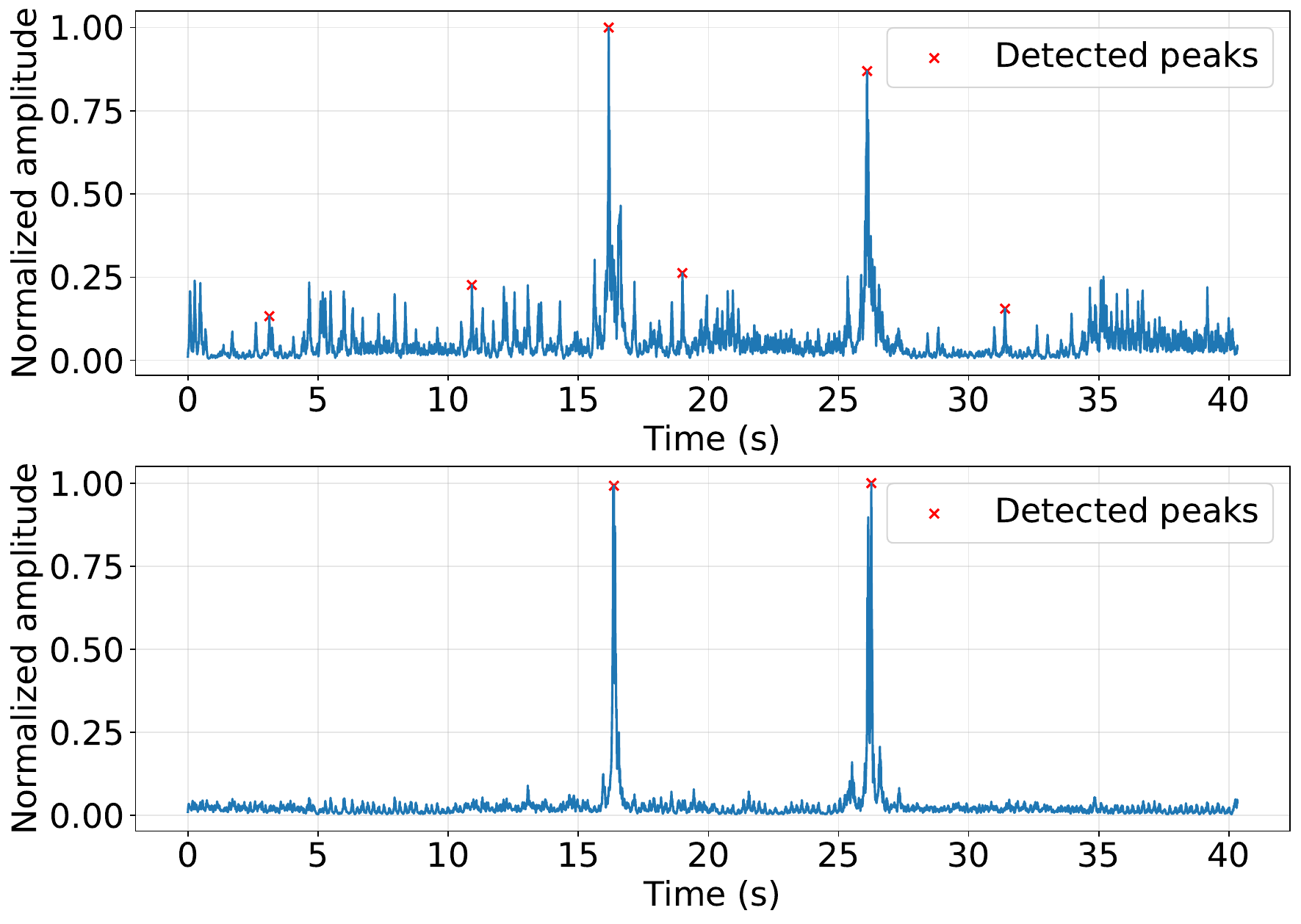}
        \caption{Detection results. (Top: cPCA, bottom: aPCA)}
        \label{fig:cwt_adaptive1_2}
    \end{subfigure}
    \caption{CWT scalograms and detection results for the two D crossings scenario.
    }
    \label{fig:two_crossings}
\end{figure}

This scenario corresponds to the experiment in which the volunteer crosses the DP between the transmitter and the receiver twice. The selections of principal components and their associated scores $s_k$ for cPCA and aPCA are listed in Table~\ref{tab:pca_scores_two_crossing}, and the corresponding CWT scalograms are shown in Fig.~\ref{fig:cwt_baseline1} and Fig.~\ref{fig:cwt_adaptive1_1}. 

\begin{table}[htb!]
\centering
\renewcommand{\arraystretch}{1.2}
\setlength{\tabcolsep}{8pt}
\footnotesize
\begin{tabular}{cccccc}  
\toprule
 \multicolumn{2}{c}{cPCA} & &
\multicolumn{2}{c}{aPCA} \\ 
\cmidrule{1-2} \cmidrule{4-5}
{\centering Components ($k$)} & {Score $s_k$} & & {Components ($k$)} & {Score $s_k$}  \\
\midrule
2 & 0.560 && 2 & 0.560 \\
3 & 0.327 && 9 & 0.537 \\
4 & 0.288 && 14 & 0.528 \\
5 & 0.290 && 24 & 0.478 \\
6 & 0.336 && 12 & 0.470 \\
\midrule
Mean & 0.360 && Mean & 0.515\\
\bottomrule
\end{tabular}
\caption{Comparison between the cPCA and aPCA components selections for the two crossings scenario.}
\label{tab:pca_scores_two_crossing}
\end{table}

It can be seen that although the conventional approach successfully identifies the two DP crossing events at the expected time instants, the resulting energy ridges appear blurred and diffuse, indicating the inclusion of noise-dominated components and lower SNR. In contrast, the adaptive framework produces clearly defined vertical energy ridges that explicitly highlight the motion-induced events, demonstrating improved separation between motion-related dynamics and background noise.
Additionally, the detection results obtained using \eqref{detection} for cPCA and aPCA are shown in Fig.~\ref{fig:cwt_adaptive1_2}. One can see that the cPCA produces several false detections, whereas the aPCA correctly identifies the crossings without false detection. This is because the former has a significantly higher noise floor than the latter as can be seen in the figure.

\subsubsection{Four DP crossings (S1)}\label{s1_3}
\begin{figure}[htb!]
    \centering
    \begin{subfigure}{0.48\linewidth}
        \centering
        \includegraphics[width=\linewidth]{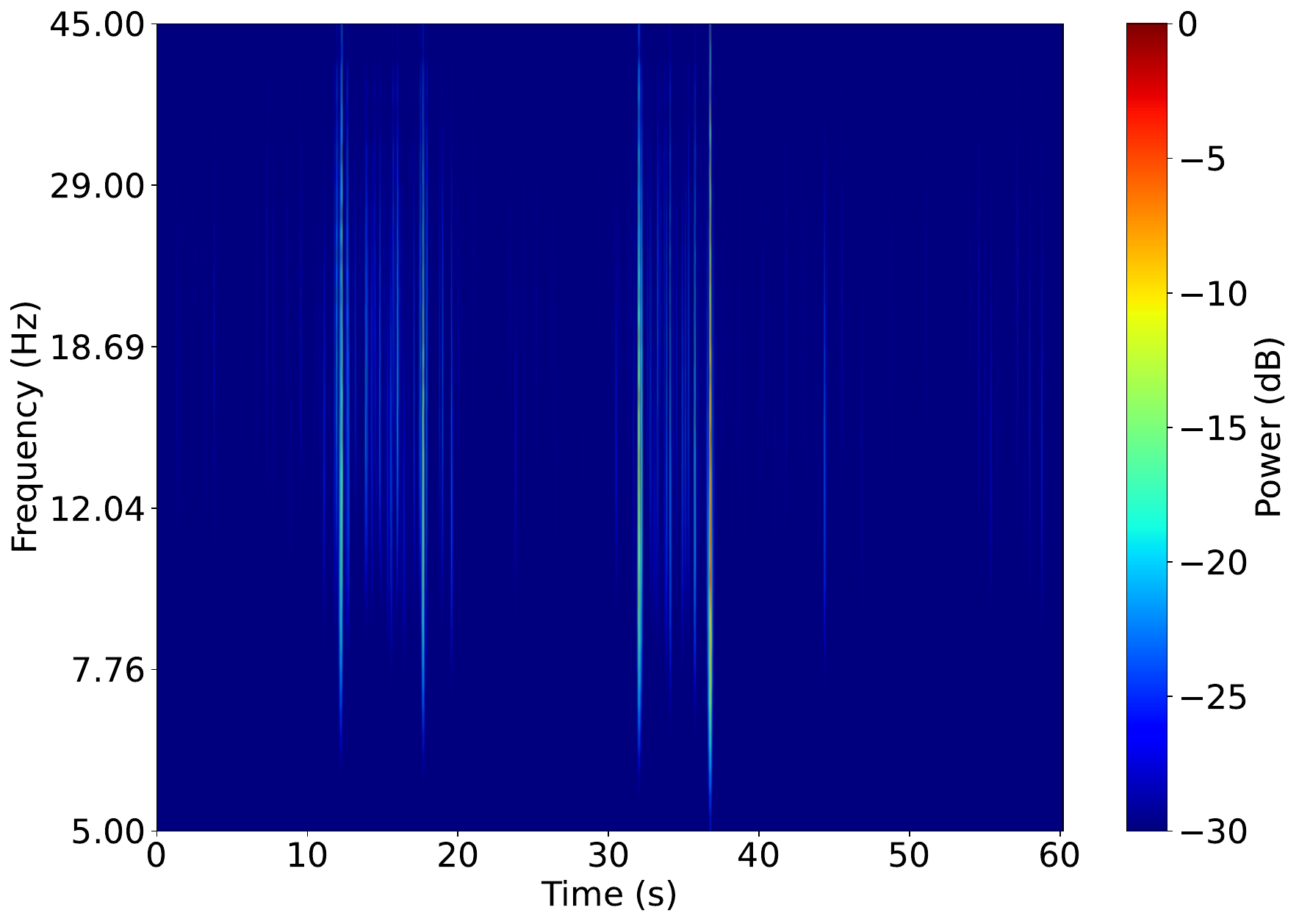}
        \caption{cPCA}
        \label{fig:cwt_baseline2_four_crossings}
    \end{subfigure}
    \begin{subfigure}{0.48\linewidth}
        \centering
        \includegraphics[width=\linewidth]{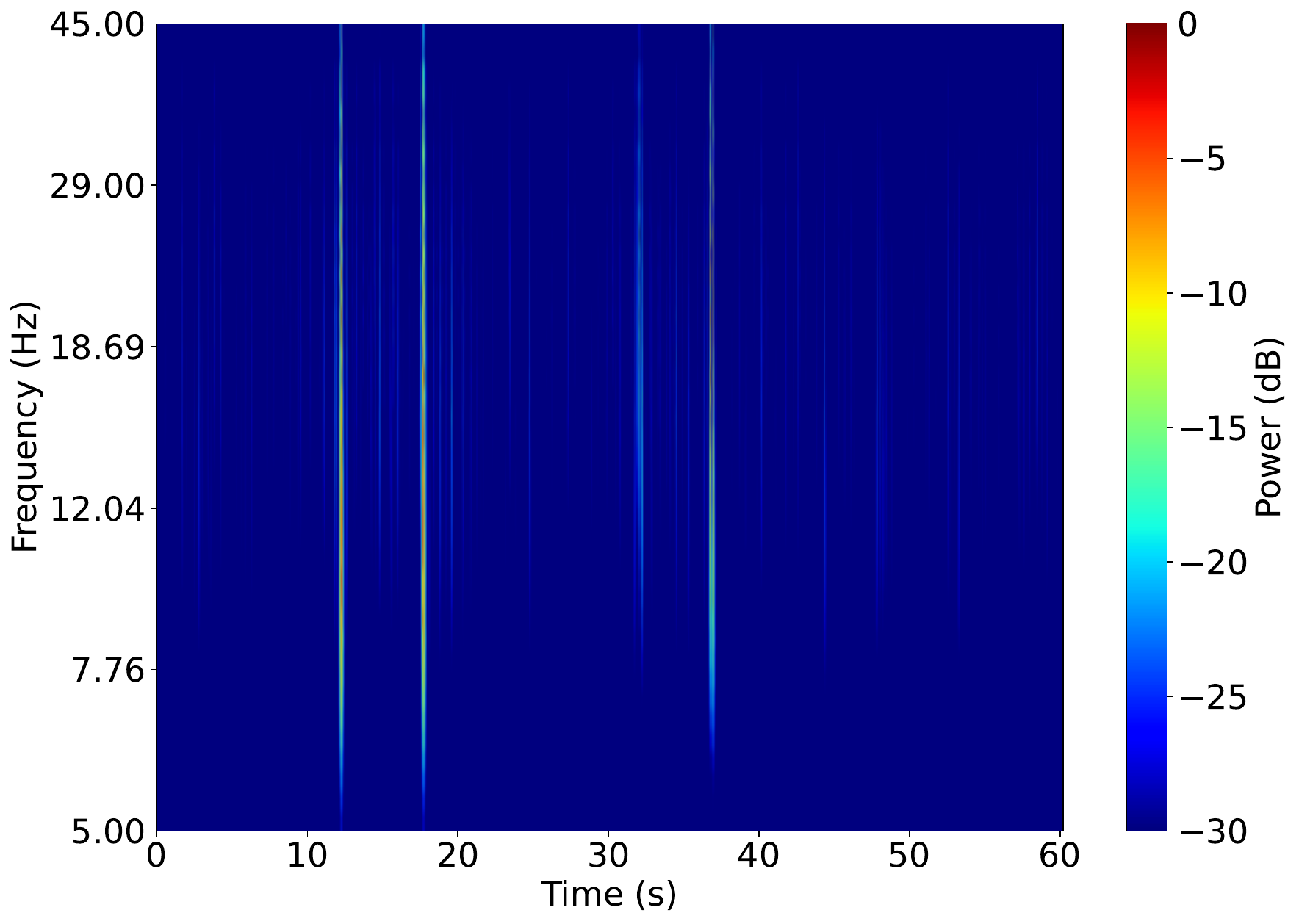}
        \caption{aPCA}
        \label{fig:cwt_adaptive2_four_crossings_1}
    \end{subfigure}\\[0.2in]
        \begin{subfigure}{0.48\linewidth}
        \centering
        \includegraphics[width=\linewidth]{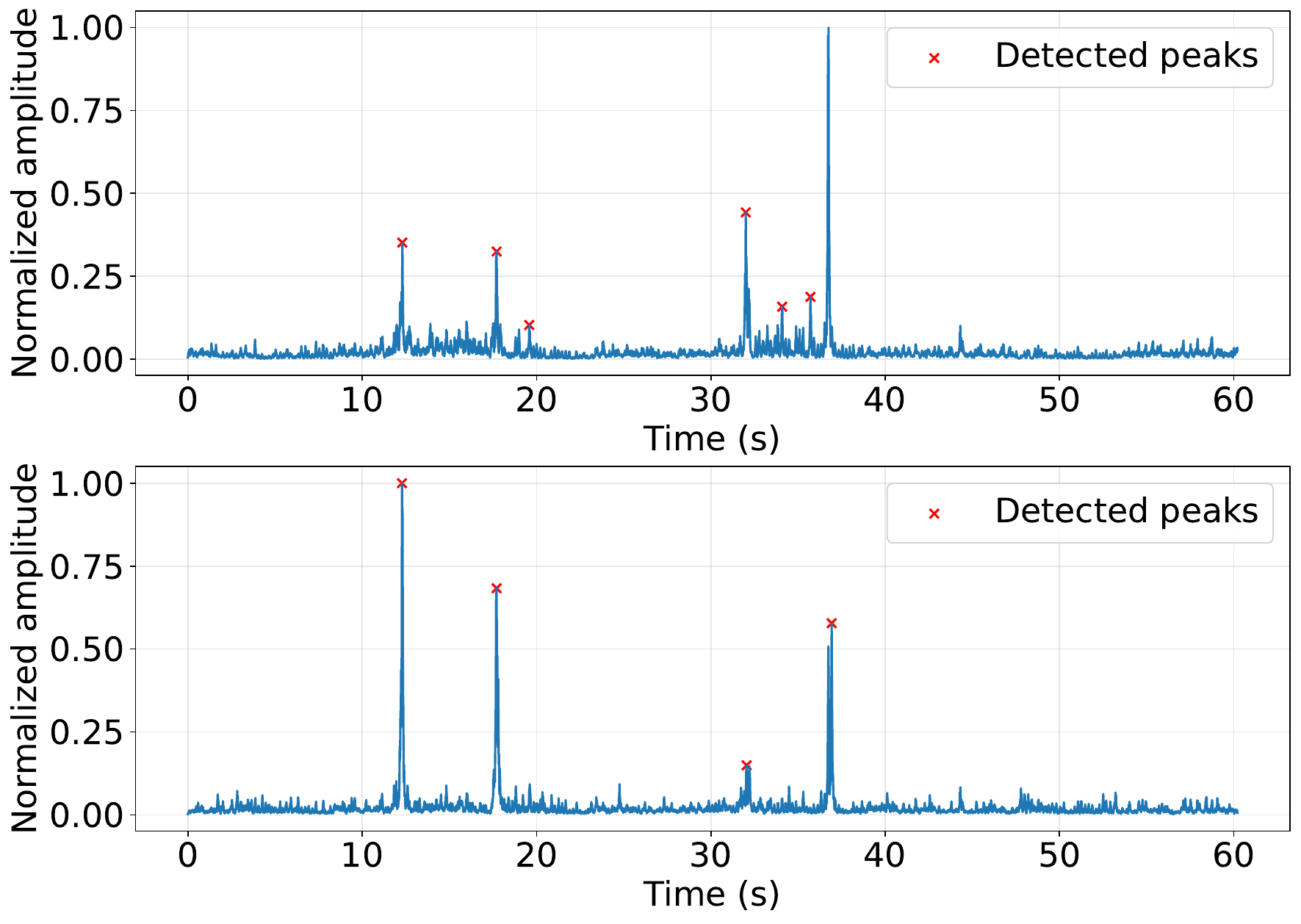}
        \caption{Detection results. (Top: cPCA, bottom: aPCA)}
        \label{fig:cwt_adaptive2_four_crossings_2}
    \end{subfigure}
    \caption{CWT scalograms and detection results for the two DP crossings scenario for the four DP crossings scenario.}
    \label{fig:four_crossings}
\end{figure}

In this scenario, the volunteer crossed the DP four times within the observation window, resulting in repeated motion-induced perturbations. The selections of principal components and their associated score $s_k$ for cPCA and aPCA are listed in Table~\ref{tab:pca_scores_four_crossing}, and corresponding CWT scalograms are shown in Fig.~\ref{fig:cwt_baseline2_four_crossings} and Fig.~\ref{fig:cwt_adaptive2_four_crossings_1}. 

\begin{table}[htb!]
\centering
\renewcommand{\arraystretch}{1.2}
\setlength{\tabcolsep}{8pt}
\footnotesize
\begin{tabular}{cccccc}   
\toprule
 \multicolumn{2}{c}{cPCA} & &
\multicolumn{2}{c}{aPCA} \\ 
\cmidrule{1-2} \cmidrule{4-5}
{\centering Components ($k$)} & {Score $s_k$} & & {Components ($k$)} & {Score $s_k$}  \\
\midrule
2 & 0.596 && 2 & 0.596 \\
3 & 0.246 && 5 & 0.414 \\
4 & 0.340 && 12 & 0.368 \\
5 & 0.414 && 13 & 0.347 \\
6 & 0.234 && 15 & 0.342 \\
\midrule
Mean & 0.366 && Mean & 0.413\\
\bottomrule
\end{tabular}
\caption{Comparison between the cPCA and aPCA components selection for the four crossings scenario. }\label{tab:pca_scores_four_crossing}
\end{table}

It can be seen from the figures that the cPCA reveals four prominent vertical ridges corresponding to the expected crossings, though it also exhibits additional spurious ridges between these events. These intermediate ridges are indicative of noise-dominated components and may be mistakenly interpreted as additional crossings, leading to ambiguous detection results.  In contrast, the aPCA framework produces a sparse and well-structured scalogram consisting of three strong vertical ridges and one weaker ridge. Although one crossing appears with reduced intensity, these four ridges are the only salient features present in the scalogram, allowing them to be reliably associated with genuine motion events. This absence of spurious detections highlights the improved robustness of the aPCA selection in suppressing noise while preserving motion-induced dynamics.
The detection results are shown in Fig.~\ref{fig:cwt_adaptive1_2}. Similarly to the case of two DP crossings, cPCA produces several false detections due to the higher noise floor, whereas aPCA correctly identifies the crossings.

\subsubsection{Experiment over separate offices (S2)}\label{s2_1}
\begin{figure}[htb!]
    \centering
    \begin{subfigure}{0.48\linewidth}
        \centering
        \includegraphics[width=\linewidth]{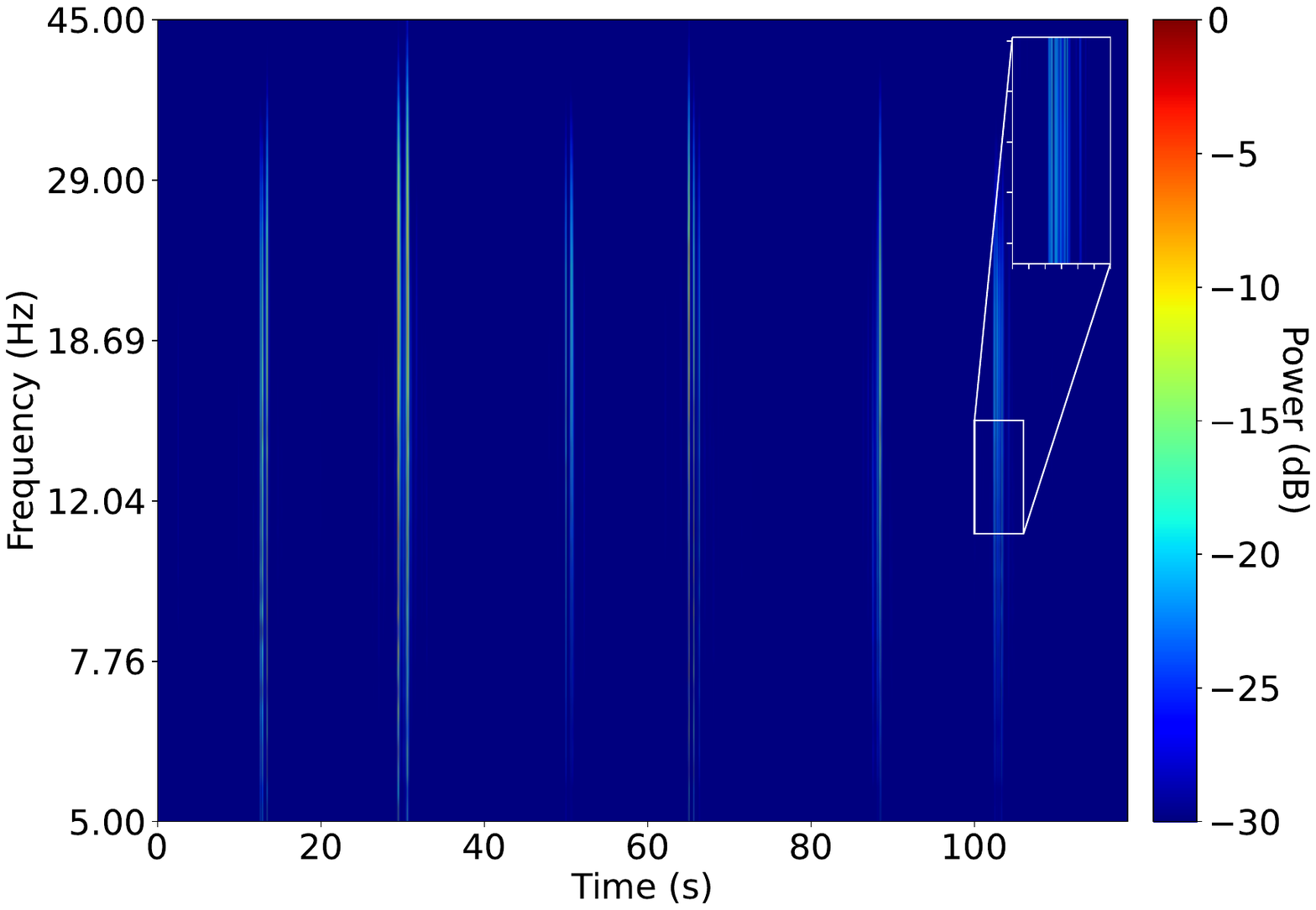}
        \caption{cPCA (PC2--PC6)}
        \label{fig:cwt_baseline2_across_offices}
    \end{subfigure}
        \hfill
    \begin{subfigure}{0.48\linewidth}
        \centering
        \includegraphics[width=\linewidth]{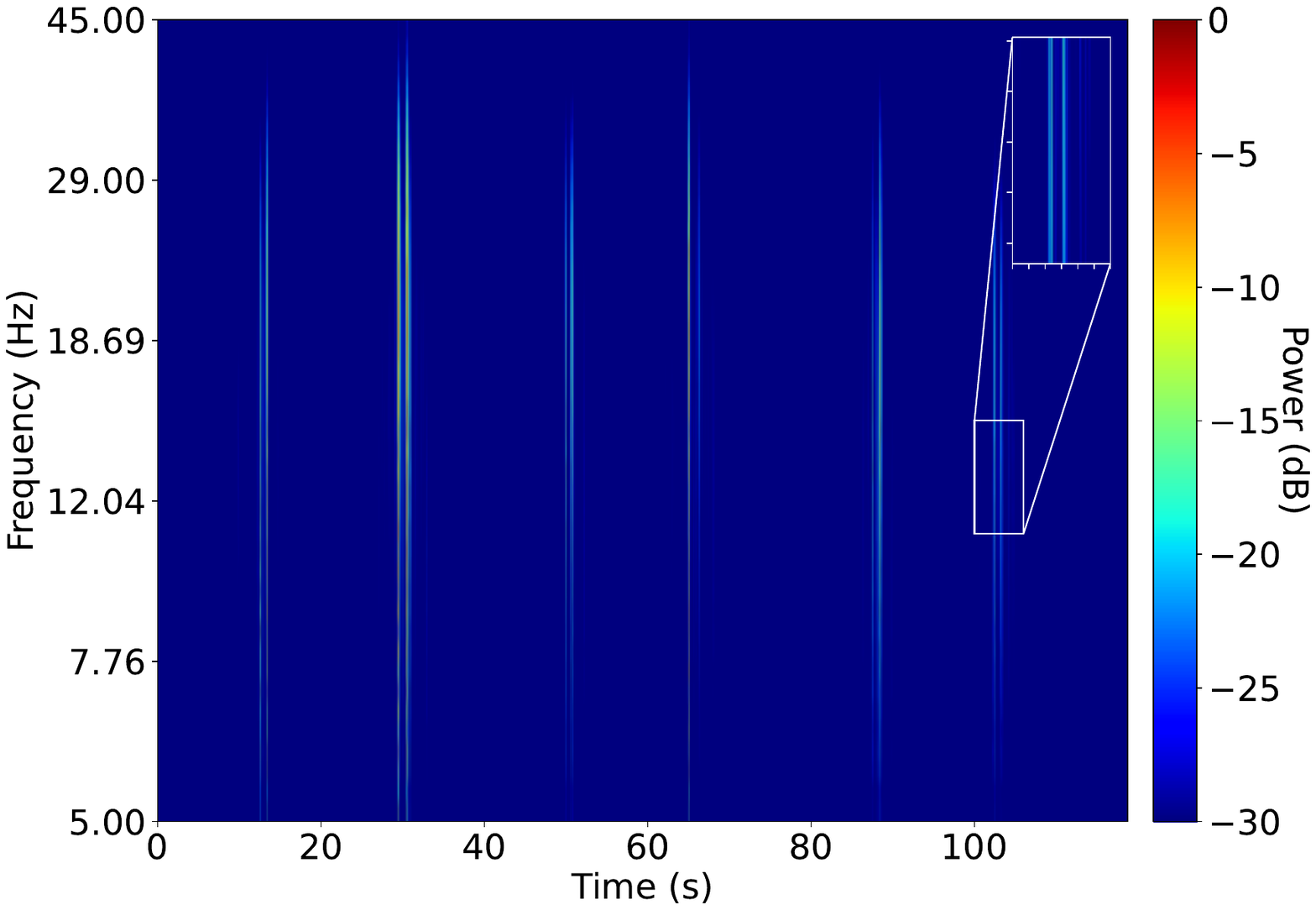}
        \caption{aPCA (proposed)}
        \label{fig:cwt_adaptive2_across_offices_1}
    \end{subfigure}\\[0.2in]
        \begin{subfigure}{0.48\linewidth}
        \centering
        \includegraphics[width=\linewidth]{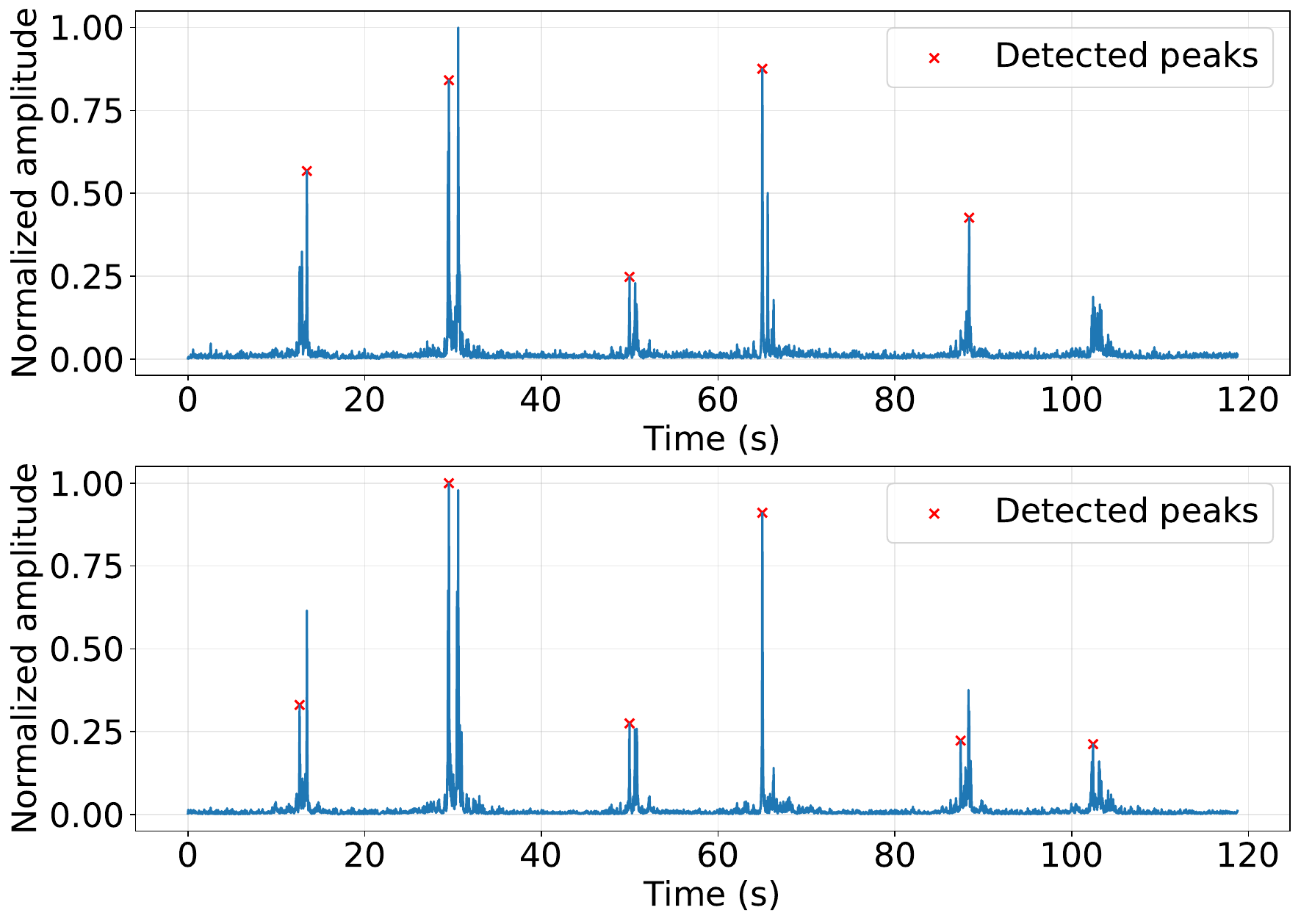}
        \caption{Detection results. (Top: cPCA, bottom: aPCA)}
        \label{fig:cwt_adaptive2_across_offices_2}
    \end{subfigure}
    \caption{CWT scalograms and detection results for the two DP crossings scenario for the across offices scenario.}
    \label{fig:across_offices}
\end{figure}

The last experiment was conducted in two offices, as shown in Fig.~\ref{fig:demo6}. The selections of principal components and their associated score $s_k$ for cPCA and aPCA are listed in Table~\ref{tab:pca_scores_across_offices}, while the CWT scalograms and detection results are shown in Fig.~\ref{fig:across_offices}.

In this scenario, the adaptive selection provides a slight improvement since most of the principal components already exhibit low noise levels. Consequently, cPCA and aPCA produce relatively clean time-frequency representations with minimal noise, and the adaptive selection provides modest improvement, indicating that most principal components are already informative in this set. Nonetheless, it should be noted that the aPCA outperforms the cPCA in terms of increasing the SNR as zoomed in Fig.~\ref{fig:cwt_baseline2_across_offices} and Fig.~\ref{fig:cwt_adaptive2_across_offices_1}, where the noise is suppressed by the aPCA and two sharp peaks are shown. These two adjacent peaks around the crossings are because the signals vary significantly when the volunteer enters and leaves the DP of the SDR and the router, which can potentially be used to identify the walking speed.

\begin{table}[htb]
\centering
\renewcommand{\arraystretch}{1.2}
\setlength{\tabcolsep}{8pt}
{\footnotesize
\begin{tabular}{cccccc}   
\toprule
 \multicolumn{2}{c}{cPCA} & &
\multicolumn{2}{c}{aPCA} \\ 
\cmidrule{1-2} \cmidrule{4-5}
{\centering Components ($k$)} & {Score $s_k$} & & {Components ($k$)} & {Score $s_k$}  \\
\midrule
2 & 0.861 && 2 & 0.861 \\
3 & 0.787 && 3 & 0.787 \\
4 & 0.625 && 8 & 0.749 \\
5 & 0.721 && 5 & 0.721 \\
6 & 0.556 && 9 & 0.709 \\
\midrule
Mean & 0.710 && Mean & 0.765 \\
\bottomrule
\end{tabular}}
\caption{Comparison between the cPCA and aPCA components selection for the across two offices scenario. }\label{tab:pca_scores_across_offices}
\end{table}

\subsection{Summary of the experiments}

Table~\ref{tab:selected_pcs} summarises the results of the experiments of the proposed aPCA selection strategy for different scenarios. 
\begin{table}[htb]
\centering
\footnotesize
\begin{tabular}{ccc}
\toprule
\multirow{2}{*}{Scenario} & \multicolumn{2}{c}{Selection of principal components}    \\ \cmidrule{2-3} 
                  & \multicolumn{1}{c}{cPCA} & aPCA  \\ \midrule
No crossings (S1)      & \multicolumn{1}{c}{2 -- 6} &  1, 5, 4, 15, 22\\
Two crossings (S1)    & \multicolumn{1}{c}{2 -- 6} &  2, 9, 14, 24, 12\\  
Four crossings (S1)     & \multicolumn{1}{c}{2 -- 6} &  2, 5, 12, 13, 15\\  
Across offices (S2)     & \multicolumn{1}{c}{2 -- 6} &  2, 3, 8, 5, 9\\  
\bottomrule
\end{tabular}
\caption{Selected principal components for cPCA and aPCA across scenarios.}
\label{tab:selected_pcs}
\end{table}

In motion-present cases (two crossings, four DP crossings and across offices), the adaptive algorithm consistently assigns higher scores to components beyond the first principal component, successfully rejecting the first component, which is often dominated by static reflections and broadband noise. In particular, the second principal component is selected in all motion scenarios, reflecting its strong association with motion-induced spectral dynamics. 

In contrast, for the no-crossing control case, where the observed signal primarily consists of background variations rather than motion-induced perturbations, the adaptive method includes the first principal component among the selected set. This highlights the adaptive nature of the proposed scoring framework, which does not impose fixed assumptions about component relevance, but instead selects components based on their spectral characteristics and consistency with the underlying physical scenario.

The robustness of the proposed aPCA framework is also evaluated by examining multiple experiments in different scenarios. As demonstrated in Table~\ref{tab:selected_pcs} and Fig.~\ref{fig:two_crossings} - Fig.\ref{fig:across_offices}, the method exhibits consistent components selection and detection behavior under varying motion conditions. In trials involving DP crossings, the adaptive scoring reliably suppresses noise-dominated components while consistently selecting components associated with motion-induced spectral structure, most notably the second principal component. Conversely, in the no-crossing control scenario, the method appropriately refrains from emphasising motion-related components and avoids spurious detections. These results demonstrate that the proposed framework adapts naturally to different propagation and motion conditions, providing stable performance across datasets without requiring manual tuning or prior knowledge of the environment.

\section{Discussion}\label{sec:discussion}

By incorporating spectral domain features, this paper proposes an adaptive PCA algorithm to select the most suitable principal components for the TWD system. Unlike conventional PCA, which selects the principal components in a fixed manner, the proposed method demonstrates improved performance in a range of scenarios. 
The experimental results confirm that the adaptive scoring mechanism effectively prioritizes motion-relevant components, leading to clearer and more localized energy structures in the wavelet domain. The consistent selection of the second principal component further supports its role in capturing motion-induced dynamics, while noise-dominated components are suppressed. At the same time, this framework remains flexible in control scenarios, avoiding rigid assumptions and reducing false detections.

However, the proposed method relies on the dominant spectral peak of the second principal component to define the reference band, which may propagate errors to the scores of all components. In addition, the proposed algorithm uses a fixed number of components, limiting its ability to fully exploit the available information. Therefore, an adaptive threshold is preferred to the optimal number of components. These will be included in future research.

\bibliographystyle{unsrt}
\bibliography{refs}

\end{document}